\def\({\left(}
\def\){\right)}
\def\[{\left[}
\def\]{\right]}
\def\a{\alpha}

\def\d{{\cal D}}

\def\g{\gamma}

\def\l{\lambda}
\def\O{\Omega}

\def\z{\zeta}
\def\bz{\bar\zeta}
\def\vect#1{\skew{-1}{\mathaccent"017E}{#1}}

\def\vp{\vect p}
\def\vq{\vect q}
\def\vx{\vect x}
\def\vy{\vect y}
\def\hvp{\hat{\bf p}}

\def\hr{\hat{r}}
\def\x{{x}}

\def\bpsi{\bar\psi}

\def\D{\hat D}
\def\K{\hat K}
\def\Q{\hat Q}
\def\q{\hat q}
\def\p{\hat p}

\def\S{\hat S}

\def\JPA{J. Phys. A \,}
\def\PRL{Phys. Rev. Lett. \,}
\def\JMP{J. of Mod. Phys. \,}
\def\be{\begin{equation}}
\def\ee{\end{equation}}
\def\exp#1{{\rm exp}\left\{#1\rlap{\phantom{\big)}}\right\}} %

\documentstyle[12pt,epsfig]{article}
\begin{document}
\title{\bf Field of homogeneous Plane in Quantum Electrodynamics\\}
\author
{ $\mbox{{\bf I.V. Fialkovsky}}^1,\ \mbox{{\bf V.N. Markov}}^2,\
\mbox{{\bf Yu.M. Pis'mak}}^{1,3}$\\
\\
 {\normalsize
 Department of Theoretical Physics,  State University Saint-Petersburg, $
 \mbox{Russia}^1$
 }\\
{\normalsize Department of Theoretical Physics, Saint-Petersburg
Nuclear Physics Institute, $\mbox{ Russia}^2$}\\ { \normalsize
Institute for Theoretical Physics, University Heidelberg,
$\mbox{Germany}^3$} } \maketitle

\begin{abstract}
  We study quantum electrodynamics coupled to the matter field on singular background, which we call
defect.  For defect on the infinite plane we calculated the fermion
propagator and mean electromagnetic field. We show that at large
distances from the defect plane, the  electromagnetic field is
constant what is in agreement with the classical results. The
quantum corrections determining the field near the plane are
calculated in the leading order of perturbation theory.
\end{abstract}

\section{Introduction}
It is a long-standing problem to formulate a quantum gauge theory
with boundaries \cite{Deutsch,Candelas,Kay}. There are different approaches for its
solution.  One can impose boundary condition on the gauge and
ghost fields \cite{8}. Other possibility in quantum
electrodynamics (QED) is to put ``physical" boundary conditions on
the electromagnetic field strength and use either explicitly
gauge-invariant methods (see for example \cite{Schwinger}), or
consider these boundary condition as constrains quantizing the
potential $A_\mu$. From the formal point of view, shortcoming of
the models of such a kind is that the gauge invariance,
renormalizability or locality appear to be broken. There is
another reason to doubt that these models describe correctly the
physics of interaction of quantum field with the boundary, since
the same sharp condition is imposed on all the modes of the
fields. In reality the modes of the field with high enough
frequencies  are not constrained by the matter and are not
influenced by its presence. We restrict ourself with these short
comments about the method of quantization with boundary constrains
on the fields because it is not used in this paper\footnote{Many
results of application of this method for investigation of Casimir
effect are presented in \cite{Bordag,Milton}}.

    The other approach which we follow here is to model boundary
effects with a static background (defect) coupled to the quantum
fields. The simplest background is the singular one. The first
attempt to build renormalizable local QFT with boundaries in this
way has been made by Symanzik in \cite{Sim}. It was shown that
some simple boundary conditions (namely Dirichlet and Neumann
ones) are implemented by adding a singular interaction to the
action density. The technique was generalized to manifolds with
curved boundaries in subsequent work \cite{Osborn}. Recently the
models of the scalar field theory in 2- and 3-dimensions were
considered in \cite{Jaffe1,Jaffe2,Jaffe3} (in the later paper
structure of divergencies and corresponding renormalization
procedure are also discussed). The method was successfully applied
for the investigation of the critical behavior of magnets and
alloys with free surfaces \cite{Diehl1,Diehl2}. In context of the
Casimir problem the models of interaction of the free quantum
field with singular background ($\delta$ function-like potential)
were considered by many authors \cite{free,prop,prop 1,free 4,free 5,free 8}. Many important results
are known also for non-relativistic quantum mechanic systems with
$\delta$-potential \cite{qm,qm 1}.

The problem of renormalization of QED with singular background
ware considered in \cite{QED,QEDa}. The radiative QED correction
to the Casimir energy was calculated in \cite{QEDb}. A review on
Casimir effect was published recently \cite{Milton}, where most
results and controversies in the field are discussed.

    Theory  of Casimir effect based on assumption  that it is a
macroscopic phenomena generating by vacuum fluctuations of QED
fields is in a good agreement with experimental data \cite{Lam,Bressi,Sukenik}. It
is naturally to expect that the Casimir effect is not unique
macro-manifestation of quantum fields' fluctuations, and in many
situations the description of macro-system behavior obtained by
classical electrodynamics (ED) needs essential quantum
corrections. The calculation of them is a theoretical task being
of practical importance for development of nano-technologies and
design of microdevices \cite{Chan}. A possible method of its
solution is proposed in this paper.

    A typical ED statement of problem is to find the electric and
magnetic fields for given boundary conditions, charge and current
distributions \cite{Landau,Jackson}. In this paper we consider
the QED version of simplest problem of such a kind. We study the
gauge invariant, local, renormalizable model of simple defect on
the plane in QED. It is suggested for calculation of quantum
corrections for the fields of a plane in classical
electrodynamics. It is essential that for renormalizability of the
model  the direct interactions of the boundary both with the
photon and with the Dirac's fields are necessary. We calculate the
leading order approximations for mean strengths of electric and
magnetic fields expressed by usual relations in terms of
components of electromagnetic field tensor. The results look like
classical ones on the large distance from the plane. For small
distances $r\rightarrow 0$ the strength of fields appear to be
singular as  $C_1/r^{2}+C_2/r$, where $C_1$, $C_2$ are constants.
It is essentially non-classical effect generated by interaction of
the Dirac fields with defect on the plane. In the framework of the
free scalar field theory with singular background a similar
phenomena was studied in \cite{dirac16}.

\section{Statement of the problem}

We consider the QED with homogeneous defect on the infinite plane
$x_3=0$ invariant in respect to coordinate reflections. It is
specified by the action functional of the form
   \be
   S_{def}
   (\bpsi,\psi, A; \lambda,q, l) =  S_{\lambda q}
   (\bpsi,\psi) + S_l
   ( A),
   \label{2_1}
   \ee
   $$
   S_{\lambda q}(\bpsi ,\psi) \equiv \int \bpsi (\vx,0)(\l+ \q)
   \psi(\vx,0) d\vx , \
   \ S_l(A) \equiv\int l A(\vx,0)d\vx+
   \int l' \partial_3^2 A(\vx,0)d\vx,
   $$
   where
$\bpsi$, $\psi$ are the Dirac's spinor fields, $A$ is the
electromagnetic vector-potential,  $q$, $l,\ l'$ are fixed
4-vectors, $\q= q_\mu \gamma^\mu$ ($\gamma^\mu$ are the Dirac's
gamma-matrices), and we used the short hand notations for the
4-vector $x$: $x=(x_0,x_1,x_2,x_3)=({\vec x}, x_3)$. The notations
of this kind will also be used later. For gauge invariance of the
defect action it is necessary to set $l_3=l'_3=0$ \footnote{We
consider the gauge transformations $A_\mu \rightarrow
A_\mu+\partial_{\mu} \phi$ not changing asymptotic of the field
$A_\mu (x)$ for large $x$. It follows from this assumption that
$\lim_{x_i \to \pm \infty} \phi(x)=\phi_0$, where $\phi_0$ is a
constant being the same for all $i=0,1,2,3$. Therefore $S_l(A)$ is
gauge invariant.}.
 With this
restriction the functional (\ref{2_1}) is a most general form of
local defect action without parameters of negative dimension and
being invariant in respect to coordinate reflections and gauge
transformations. Therefore in virtue of usual criteria of
renormalization theory the model defined by addition to the QED
action the defect action (\ref{2_1}) is renormalizable \cite{Sim}.
It remains to be renormalizable for $l'=0$ too, because as we see
below, there are no divergencies which need for their cancellation
the $l'$-term in the $S_l$ term.

 The physical meaning of the vector $l=(\vec{l},0)$ is very simple.
It defines the classical 4-current on the plane defect. By
neglecting in our model the interaction of the photon and Dirac
fields, the mean electromagnetic field coincides with solution of
Maxwell equations with the current (supported on the plane)
defined by $S_l$. Vector $q$ and scalar $\lambda$ describe the
interaction of current and density of Dirac field with material
defect. Interaction of vacuum fluctuations of the Dirac field with
the background generates quantum corrections to usual classical
effects.

   We calculate the leading approximation for electromagnetic
field generated by the defect. Only the $l$-term in $S_l$ appears
to be necessary for cancellation of ultra-violet divergences in
our results. We set $l'=0$ and do not consider a trivial
contribution to the first order effects from $l'$-term in $S_l$.
We choose the vector $\vec {l}$ proportional to $\vec q$: ${\vec
l}={\vec q} \xi $, and show that this ``minimal" form of $S_l$
with only one extra to $S_f$ parameter $\xi$ provides the
cancellation of divergencies by renormalization of $\xi$.

    The full action of our model has the form
 \be
 S(\bpsi, \psi, A)=
    S_{QED}(\bpsi ,\psi,A)+ S_{def}(\bpsi ,\psi, A;\l, q, \xi),
\label{2_2}
  \ee
where $S_{QED}(\bpsi ,\psi,A)$ is the usual QED action: $$
 S_{QED}(\bpsi ,\psi,A) =
 \int \bpsi (x)(i\hat{\partial}
 -e\hat{A}(x) - m )\psi(x)dx
 -\frac14\int F_{\mu\nu}(x)F^{\mu\nu}(x)dx.
$$ Here, $$ F_{\mu\nu}=\partial_\mu A_\nu-\partial_\nu A_\mu,\ \
    \hat{A}=A_\mu \g^\mu,
$$ $A_\mu$ is the potential of electromagnetic fields and gamma
matrices fulfill the commutation relations $$
\{\g_\mu,\g_\nu\}\equiv \g_\mu\g_\nu +\g_\nu\g_\mu
    = 2g_{\mu\nu}
$$
with metric tensor $g_{\mu\nu}=\mbox{diag}\{1,-1,-1,-1\}$. The
considered model is invariant in respect to translations and
Lorenz transformations leaving unchanged the coordinate $x_3$, if
the vectors $\vec {q}$ transformed correspondingly.

In virtue of Lorenz-invariance of the model in the sense mentioned
above, we can classify the defect properties with the value of the
invariant $\vq^2\equiv q_0^2-q_1^2-q_2^2$. So, we have 3 cases:
$\vq^2=\kappa^2>0$, $\vq^2=-\kappa^2<0$ and $\vq^2=0$, and we can
choose (and restrict ourselves with) three coordinate systems
where $\vq=(\kappa,0,0)$, $\vq=(0,\kappa,0)$ ¨
$\vq=(\kappa,\kappa,0)$ accordingly.

This means that interaction of QED fields with the defect is
described by four parameters: $\kappa$, $\tau \equiv q_3$,
$\lambda$ and $\xi$.

In this paper we calculate the mean tensor $ F_{\mu\nu}$ of
electromagnetic field:
    \be {\cal F}_{\mu\nu}=C\int F_{\mu\nu} e^{i
S(\bpsi, \psi, A)} DA D\bpsi D\psi, \label{2_3}
    \ee
where the constant $C$ is defined as follows
  $$ C^{-1}= \int e^{i S(\bpsi, \psi, A)} DA
D\bpsi D\psi.
   $$
    For calculations, it is more convenient to use the Euclidean
version of the model. It is obtained by the usual Wick rotation
  $$
t\rightarrow -it,\quad A_0\rightarrow iA_0,\quad \g_0\rightarrow
i\g_0, \quad q_0 \rightarrow iq_0.
  $$
Then the Euclidean action can be  written as
  $$
 S^{E} = S_{QED}^{E} + S_{def}^{E},
  $$
where
  $$
S_{QED}^{E}  = \bpsi (i\hat{\partial} -e\hat{A} + m )\psi
+\frac{1}{4}F_{\mu\nu}F_{\mu\nu}, \ \ S_{def}^E(\bpsi ,\psi, A;
\lambda,q, \xi)= \bar \Psi(\l+ \q) \Psi + \xi ({\vec q}{\vec A}),
  $$
  $$
  {\bar \Psi}(\vx)=\bpsi (\vx,0), \qquad
  {\Psi}(\vx)=\psi (\vx,0).
  $$
The matrix $\hat{a}$ corresponding to 4-vector $a$ is defined in
Euclidean theory as $\hat{a}=\sum_{\mu=0}^3 a_\mu\g_\mu$. For
convenience of calculations we shall use the projection operator
onto the plane $x_3=0$
\be \O(\vx , y)= \delta(\vx-\vy)\delta(y_3) \label{omega}
  \ee
and present the function ${\bar \Psi}(\vx)$, ${\Psi}(\vx)$ in the
form
    $$
    \bar\Psi(\vx) = \int \bpsi(y)\O^T(y,\vx)dy\equiv \bpsi
    \O^T, \quad
    \Psi(\vx) = \int \O(\vx , y)\psi(y)dy\equiv  \O\psi.
    $$
The tensor $ F_{\mu\nu}$ is gauge invariant, therefore ${\cal
F}_{\mu\nu}$ is independent of the choose of the gauge. We provide
calculations in the Feynman's gauge%
\footnote{This gauge can be fixed by the usual Faddeev-Popov
trick}, using the formula $$
   D_{\mu\nu}(x,y)= \frac{\delta_{\mu\nu}}{4\pi^2 (x-y)^2}
$$ for the photon propagator in configuration space. Integrating
by parts in functional integral (\ref{2_3}) we obtain
  \be
{\cal F}_{\mu\nu}(x)= \frac{1}{2\pi^2}\int d^4y \frac{
(x_{\nu}-y_\nu)J_\mu(y)- (x_{\mu}-y_\mu)J_\nu(y)}{(x-y)^4},
\label{2_4}
  \ee
where
   \be J_\mu(y)=ej_\mu(y)-l_\mu\delta(y_3), \
  j_\mu(y)=C\int e^{-S(\bpsi ,\psi,\vq)}
  \bpsi(y)\gamma_\mu\psi(y) D\bpsi D\psi . \label{2_5}
   \ee

In virtue of invariance  of the action (\ref{2_2}) in respect to
translation of coordinates $x_0, x_1, x_2$ and reflection of
$x_3$, $j_\mu(y)$ is an even function of coordinate $y_3$ only:
$J(y)=J(y_3)= J(-y_3)$, and after integration over $\vec{y}$ in
(\ref{2_4}) we obtain the following result
  \be
{\cal F}_{\mu\nu}(x)= \frac{1}{2}\int_{-x_3}^{x_3} [\delta_{\nu 3}
J_\mu(y_3) - \delta_{\mu 3} J_\nu(y_3)] dy_3 =
   \label{2_6}
  \ee
$$ =\mbox{Sign}(x_3)\left\{ e\int_{0}^{|x_3|} [\delta_{\nu 3}
j_\mu(y_3) - \delta_{\mu 3} j_\nu(y_3)] dy_3 -\xi (\delta_{\nu 3}
q_\mu - \delta_{\mu 3} q_\nu )\right\} , $$
where $\mbox{Sign}(x_3)$ is the signum-function: $\mbox{Sign}(x_3)=
x_3/|x_3|$.

    The vector $j_\mu$ in (\ref{2_5}) is the current generated by
vacuum fluctuations of Dirac fields. In virtue of Farri's theorem,
it vanishes in absence of defect. By usual methods of
renormalization theory and its modification for the quantum field
theory with singular background \cite{Sim} one can prove that in
the framework of renormalized perturbation theory $j_{\mu}(y_3)$
is presented by a sum of diagrams with all the necessary
subtractions. Therefore it is finite (for the leading
approximation it will be clear from the evident formula below),
and in calculation of (\ref{2_6}) there is the only problem - one
with non-integrable singularity of $j_\mu (y_3)$ at $y_3=0$.
Therefore the integral (\ref{2_6}) needs a regularization. Since
for $x_3 \neq 0$ the derivative of ${\cal F}_{\mu\nu}(x_3)$ in
respect to $x_3$ is finite, we obtain the finite value of ${\cal
F}_{\mu\nu}(x_3)$ subtracting  from integral in the left hand of
(\ref{2_6}) a constant dependent on the chosen regularization.
This subtraction can be generated by term $S_l$ with appropriate
choice of the parameter $\xi$. Therefore the $l'$-term in $S_l$ is
not necessary for renormalizability of considered model.

 The leading approximation $j^{(0)}_\mu$ for $j_{\mu}$ can be
presented as
   \be
 j_{\mu}^{(0)}\equiv  Tr ( \S(y,y)\gamma_\mu).
   \label{2_7}
   \ee
Here $\S(x,y)$ is the free fermion propagator in the theory with
defect $S_{\lambda q}(\bpsi,\psi)$
  \be
  \S_{\alpha\beta} (x,y)=C_f\int e^{-S_f(\bpsi
  ,\psi, q)} \bpsi_\beta(y) \psi_\alpha(x) D\bpsi D\psi.
  \label{2_8}
   \ee
We used in (\ref{2_8}) the notation $S_f(\bpsi ,\psi,q)$ for the
fermion part of the  action $S^E$
$$
S_f(\bpsi ,\psi, q) = S^E(\bpsi ,\psi,A;\l, q,\xi)|_{A=0}= \bpsi
\K\psi + \bar \Psi(\l+ \q)  \Psi, \ \K\equiv i\hat{\partial} +m,
$$ and $$ C_f^{-1}= \int e^{-S_f(\bpsi, \psi,q)} D\bpsi D\psi.
$$
We restrict ourselves with calculation of the main approximation
for ${\cal F_{\mu\nu}}$, and in virtue of (\ref{2_7}), the most
nontrivial part of this problem is to find an evident form for the
fermion propagator $\S$ defined by (\ref{2_8}).

\section{Calculation of $\S$}
The right hand side of (\ref{2_8}) can be obtained by
differentiation of  generating functional
$$
G({\bar{\eta}}, \eta)
\equiv C_f \int \exp{-S_f(\bpsi ,\psi;q)+ \bpsi \eta+
\bar{\eta}\psi}\d\bpsi \d\psi
$$
where $\bar{\eta}$, $\eta$ are
the fermion  sources. Obviously,
$$
\S_{\alpha\beta}
(x,y)=\frac{\delta}{\delta \bar{\eta}_\alpha(x)}
\frac{\delta}{\delta \eta_\beta(y)} G({\bar{\eta}}, \eta)|_{
\bar{\eta}, \eta =0 }, \ G(\eta,\bar\eta) =
\exp{\bar{\eta}\S\eta}.
$$
To calculate $G({\bar{\eta},\eta})$ we present the contribution of
the defect in the following form
$$
\exp{- \bar\Psi
(\l+\q) \Psi}=
    c\int \exp{ \bz\z + \bz (\l+\q)\Psi + \bar\Psi\z}
\d\bz \d\z,\ c^{-1}= \int \exp{ \bz\z } \d\bz \d\z,
$$
and write
$G({\bar{\eta},\eta})$ as $$ G({\bar{\eta},\eta})=cC_f\int \exp{-
    \bpsi\K\psi+\bz\z+
    \bpsi\eta+ \bar{\eta}\psi + \bar\Psi\z +
    \bz(\l+ \q)\Psi} \d\bpsi \d\psi D\bz D\z
$$
$$ =cC_f\int \exp{-\bpsi\K\psi+\bz \z + \bpsi\theta +
\bar{\theta}\psi}
    \d\bpsi \d\psi \d\bz \d\z,
$$
where
$$ \bar{\theta}\equiv
 \bar{\eta}+\bz(\l+\q) \O ,\quad
    \theta \equiv \eta + \O^T\z.
$$
Now, integrating over the fields $\bpsi$, $ \psi$ and then over
$\z$, $\bz$  we obtain
 $$ G(\eta,\bar\eta) =
\exp{\bar{\eta}\S\eta}, $$
where
\be
  \S = \D+\S_{def},\qquad \S_{def}=-\D\O^T\Q^{-1}(\l+\q) \O\D.
 \label{3_1}
 \ee
We used here the following notations
$$ \D\equiv \hat{K}^{-1}, \
\Q\equiv {\bf 1} + (\l +\q)\O\D\O^T.
$$
We see that the propagator $\S$ in our model is the sum of
propagator $\D$ of usual QED and $\S_{def}$ generated by the
defect. The propagator $\D$ does not make contribution into
(\ref{2_7}), and we can present $j^{(0)}_\mu$ as
\be
 j^{(0)}_\mu(y)=Tr (\S_{def}(y,y)\gamma_\mu).
\label{3_2} \ee

   Now, we calculate $\S_{def}$ in an evident form, which will be
used to obtain the  final result for ${\cal F}_{\mu\nu}$ in
considered approximation. In virtue of translation invariance for
coordinates ${\vec x}$, it is natural in our model to use for
calculations the 3-dimensional Fourier transformation defined as
follows
$$
F(\vx, x_3)=\frac{1}{(2\pi)^{3}}\int  d\vp\, e^{i\vp\vx}F(\vp,
x_3).
$$
For the propagator $\D$ we have $$ \D (x)\equiv\D(\vx,
x_3) =\frac{1}{(2\pi)^{3}}
    \int\frac{m+\hvp+iE\g_3 \mbox{Sign}(x_3)}{2E}\exp{-E|x_3|+i\vp\vx}d\vp,
$$
 where
$$ \hvp\equiv \vp\vec{\g},\quad E\equiv\sqrt{\vp^2+m^2}. $$
Hence,
 $$
\D(\vp, x_3)=
    \frac{m+\hvp+iE\g_3 \mbox{Sign}(x_3)}{2E}\,e^{-E|x_3|}.
    $$
For calculation of $\S_{def}(x,y)$ we note that in virtue of
(\ref{omega})
$$    \ \O\D\O^T(\vp)=
    \D(\vp,0)=
    \frac{m+\hvp }{2E}.
$$
So, we can  write
 \be
\S_{def}(x,y) =\frac{1}{(2\pi)^3} \int d\vp ~e^{i\vp(\vx-\vy)}
\S_{    def}(\vp;x_3,y_3), \label{3_3}
 \ee
 where
 $$
\S_{def}(\vp;x_3,y_3)\equiv
    -\D(\vp;x_3)\(1 + (\l+\q)\frac{m+\hvp }{2E}\)^{-1}(\l+\q)\D(\vp;-y_3)
 $$
 \be
=-\frac{[m+\p_1][M +\hr][m+\p_2]}
{2E^{2}[E\left(4+q^2+\l^2\right)+4\left(\l m -\vp \vq\right)]
}\,e^{-E(|x_3|+|y_3|)}. \label{3_4}
 \ee
  Here, we have denoted
  $$
p_1=(\vp,i\mbox{Sign}(x_3)E),\  \ \ p_2=(\vp,-i\mbox{Sign}(y_3)E),
  $$
  $$
  M=2 E\l +(\l^2+q^2)m,\ \ \ r=\( 2E\vq-(\l^2+q^2)\vp, 2Eq_3\).
  $$
The Dirac field propagator $\S$ for the defect with
$\l=q_1=q_2=q_3=0$ was calculated in \cite{prop,prop 1}, and our
result (\ref{3_4}) generalize one obtained there, and coincides
with it for particular values of defect parameters.

 Now, we use (\ref{3_4}) for calculation of electromagnetic field
generated by the defect.

\section{Calculations of ${\cal F}_{\mu\nu}$}

Setting the right hand side of (\ref{3_3}) into (\ref{3_2}) and
using (\ref{3_4}), one obtains
$$
j^{(0)} (x)=
(\vec{j}^{(0)}(x),j^{(0)}_3(x))= \frac{e}{(2\pi)^3}\int d\vp\, Tr
[({\vec\gamma},\gamma_3) S_{def}(\vp, x_3,x_3)]
$$
$$
=-\frac{e}{\pi^3}\int d\vp
\frac{e^{-2E|x_3|}}{E(E\left(4+q^2+\l^2\right)+4\left(\l m -\vp
\vq\right))}
 \Big(\vp\,(\vp\vq-m\l)-E^2\vq,~0\Big).
$$
After integration over angular variables $j^{(0)}(x)$ is
written in the form
  $$
  j^{(0)}(x) = e\, ({\vec q},0)\, \phi (x_3),
  $$
  \be
   \phi(x_3)\equiv
   \frac{1}{2\pi^2 |\vq|\alpha^2}
   \int_0^\infty  p \,e^{-2E|x_3|}
   \[2 p \alpha -(E(1-\alpha^2) +m \beta))\,
   \ln\frac{E +\a p+\beta m}{E-\alpha p +\beta m}\]
    dp \, ,
\label{4_1}
   \ee
    where
    \be
\a=\frac{4|\vq|}{4+q^2+\l^2}, \ \ \ \, \beta=\frac{4\l
}{4+q^2+\l^2}. \label{4_2}
    \ee

  For $x_3\neq 0$ the current $j^{(0)}(x_3)$ is finite  since the
integral (\ref{4_1}) converges. However we can not set $j(y_3)
=j^{(0)}(y_3)$ in (\ref{2_6}) directly, since $j^{(0)}(y_3)$ has a
non-integrable singularity at $y_3=0$. The regularized form of the
leading approximation of $\cal{F}_{\mu\nu}$ can be obtained by
introducing a cut-off parameter $\Lambda$ in (\ref{4_1}) and
substituting then of regularized current $j^{(0)}(\Lambda)$ into
(\ref{2_6}).
 In doing so one obtains the following result
      $$ {\cal F}_{\mu\nu}(x)=
   e(q_{\mu}\delta_{\nu3}-q_{\nu}\delta_{3\mu})\mbox{Sign}(x_3)
    \[ F(x_3,\Lambda)-\frac{\xi}{2}\],
   $$
   where
   $$
   F(x_3,\Lambda)\equiv
   \frac{1}{4\pi^2 |\vq|\alpha^2}
   \int_0^\Lambda dp\frac{\left(1- e^{-2E|x_3|}\right)p}{E}
   \[2 p \alpha -(E(1-\alpha^2) +m \beta)\,
   \ln\frac{E +\a p+\beta m}{E-\alpha p +\beta m}\].
    $$
    The limit ${\displaystyle \lim_{x\rightarrow \infty}} F(x,\Lambda)\equiv
F(\infty,\Lambda)$ can be calculated in an evident form
$$
F(\infty,\Lambda)=\frac{1}{4\pi^2|{\vec q}|\alpha^2 }\Bigg\{
\alpha \Lambda (\sqrt{\Lambda^2+m^2}+\beta m) + m^2[ \xi (\Lambda,
a_1, b_1)+\xi (\Lambda, a_2, b_2)]
$$
$$ +
\left(\frac{{(\Lambda^2+m^2)}(1-{{\alpha }^2})}{2}+ \beta m
\sqrt{\Lambda^2+m^2} \right) \ln
\left[\frac{\sqrt{\Lambda^2+m^2}-\alpha \Lambda  + \beta m
}{\sqrt{\Lambda^2+m^2}+\alpha \Lambda +\beta m }\right] \Bigg\}.
$$
Here
$$ \xi (\Lambda, a,b)= \frac{b }{ {\sqrt{{a^2}-1}}}  \ln
\left[ \frac{a\sqrt{\Lambda^2+m^2} -m +\Lambda \sqrt{{a^2}-1}}{a\,
m-\sqrt{\Lambda^2+m^2}} \right],
$$
and
$$ {a_1}=-\frac{ \beta
+\alpha  {\sqrt{{{\alpha }^2}+{{\beta }^2}-1}}}{1-{{\alpha }^2}},
\ \ {a_2}=-\frac{ \beta -\alpha {\sqrt{{{\alpha }^2}+{{\beta
}^2}-1}}}{1-{{\alpha }^2}},
$$
   \be
 {b_1}=\frac{({{\alpha }^2}+{{\beta }^2}) \big(\alpha  \beta
 + {\sqrt{{{\alpha }^2}+{{\beta }^2}-1}}\big)}
{2(1-{{\alpha }^2})}, \ \ {b_2}=\frac{({{\alpha }^2}+{{\beta }^2})
\big(\alpha  \beta  - {\sqrt{{{\alpha }^2}+{{\beta
}^2}-1}}\big)}{2(1-{{\alpha }^2})} \label{4_3}
  \ee

  The  asymptotic
of $F(\rho, \Lambda)$ for large $\Lambda$ can be written as
$$
F(\rho,\Lambda)=c_2 \frac{\Lambda^2}{m^2}+ c_1 \frac{\Lambda}{m}+
c_0+ f(\rho)+O\left(\frac{m^2}{\Lambda^2}\right),
$$
where
\be
   f(\rho)\equiv
   -\frac{1}{4\pi^2 |\vq|\alpha^2}
   \int_0^\infty dp\frac{ e^{-2E|\rho|}p}{E}
   \[2 p \alpha -(E(1-\alpha^2) +m \beta)\,\ln\frac{E +\a p+\beta m}{E-\alpha p +\beta m}\],
\label{4_4}
  \ee
  $c_{0}$, $c_{1}$, $c_{2}$ are the constants:
  $$
c_2=\frac{m^2}{8\pi^2|{\vec q}|\alpha^2 }\Bigg\{2 \alpha +
({{\alpha }^2}-1) \ln \frac{1+\alpha } {1-\alpha }\Bigg\},\
c_1=\frac{\beta m^2}{4\pi^2|{\vec q}|\alpha^2}{ \bigg\{2 \alpha
-\ln \frac{1+\alpha }{1-\alpha }\bigg\}},
$$
$$ c_0= c_2
+\frac{m^2}{4\pi^2|{\vec q}|\alpha^2}\Bigg\{ \frac{ \alpha{{\beta
}^2}}{ (1-{\alpha }^2)} +\frac{{b_1}\ln
\big[-{a_1}-{\sqrt{a_{1}^{2}-1}}\big] }{{\sqrt{a_{1}^{2}-1}}}+
\frac{{b_2}\ln \big[-{a_2}-{\sqrt{a_{2}^{2}-1}}\big]
}{{\sqrt{a_{2}^{2}-1}}}\Bigg\}
$$
where we used the notations (\ref{4_2}),(\ref{4_3}),  and
$O\left(\frac{m^2}{\Lambda^2}\right)\rightarrow 0$ for
$\Lambda\rightarrow \infty$. Requirement of ${\cal F}_{\mu\nu}$
being finite for $\Lambda \rightarrow \infty$ means that $\xi$
depends on $\Lambda$, and for large $\Lambda$ its asymptotic  has
the form
$$
\xi (\Lambda)= 2 \left( c_{2} \frac{\Lambda^2}{m^2} + c_{1}
\frac{\Lambda}{m}\right)+ \chi+
O\left(\frac{m^2}{\Lambda^2}\right).
$$
Here the parameter $\chi$ is the renormalized value of $\xi$ .

Thus, if we denote $\vec
{\cal F} = ({\cal F}_{03},{\cal F}_{13},{\cal F}_{23})$, then for
$\Lambda \rightarrow\infty$ we obtain the following result
 $$
  \vec{\cal F}(x)=e \,\mbox{Sign}(x_3)\,\vq \left(c_{0}+ f(x_{3})-\frac{\chi}{2}\right).
  $$
 The asymptotics of function $f(\rho)$ (\ref{4_4}) are of the form
 $$
f(\rho) \mathrel{\mathop=_{\rho\to\infty}}- \frac{\alpha m^2 e^{-2
m |\rho|}}{8|\q|(\pi m |\rho|)^{3/2} (1+\beta)}\,
 (1+O(1/\rho)),
 $$
 $$
\ f(\rho) \mathrel{\mathop=_{\rho\to 0}} - \frac{c_2}{2 m^2
\rho^2} - \frac{c_1}{2 m |\rho|} -c_0 + c_2 +O(\rho).
  $$

  Hence
  $$
\vec{\cal F}(x)\mathrel{\mathop=_{x_3\to\infty}}
e\,\mbox{Sign}(x_3)\,\vq\left( c_{0}-\frac{\chi}{2} - \frac{\alpha
m^2 e^{-2 m |x_3|}}{8|\q|(\pi m
|x_3|)^{3/2}(1+\beta)}\,(1+O(1/x_3))\right) ,
   $$
   $$
\vec{\cal F}(x)\mathrel{\mathop=_ {\x_3\to\ 0}}
-e\,\mbox{Sign}(x_3)\,\vq \left(\frac{c_2}{2 m^2 x_3^2}
+\frac{c_1}{2 m |x_3|} + \frac{\chi}{2} -c_2  +O(x_3)\right).
\label{5.5}
  $$

   We see, that on large distances the field generated by defect is of the same form
as the field of the plane in classical ED \cite{Landau,Jackson},
and on the small distance $x_3$ it has non-classical behavior $
\vec{\cal F}\sim e \vq c_2 /2 m^2x^2_3$.
 Let us consider 3 cases: 1) $q
=(\kappa,0,0,\tau)\equiv q^{(1)}$;  2) $q =(0,\kappa,0,\tau)\equiv
q^{(2)}$; 3) $q =(\kappa,\kappa,0,\tau)\equiv q^{(3)}$.
  If $q=q^{(1)}$, the defect generates pure electric field:
$ H_1=H_2=H_3=0, \ E_1=E_2=0, \ E_3=-i{\cal F}_{03}(-i\kappa) $.
For $ q=q^{(2)}$ the field is pure magnetic one: $ E_1=E_2=E_3=0,
\ H_1=H_3=0,\ H_2={\cal F}_{13} $. For $q=q^{(3)}$ there are
magnetic and electric fields: $E_1=E_2=H_1=H_3=0$, $E_3=H_2={\cal
F}_{13} $.

To present the obtained results for all three cases in a compact
form we introduce the following notations: $\mbox{F}_1 \equiv E_3$
for $q=q^{(1)}$, $\mbox{F}_2 \equiv H_2$ for $q=q^{(1)}$, and
$\mbox{F}_3 \equiv E_3$ for $q=q^{(3)}$. Then near the plane the
strengths of fields are
$$
\mbox{F}_i(x_3)\mathrel{\mathop=_ {\x_3\to 0}}
 \frac{ f_i}{m^2
x_3^2} + \frac{ g_i}{m |x_3|} + \frac{ (-1)^{i+1} e\chi\kappa
}{2}-2 f_i + O(x_3), $$ where, $$ f_1 = \frac{e m^2}{8
\pi^{2}\alpha_1^2 }\left[(1+\alpha_1^2)\mbox{Arctg}(\alpha_1)
-\alpha_1  \right],\ g_1=\frac{e \beta_1 m^2}{4
\pi^{2}\alpha_1^{2}}\left[\mbox{Arctg} (\alpha_1)-\alpha_1
\right], $$
 $$
  f_2= \frac{e m^2}{16  \pi^{2} \alpha_2^{2}}\left[
(1-\alpha_2^{2})\mbox{ln}\frac{1+\alpha_2}{1-\alpha_2}- 2\alpha_2
\right], \ g_2=\frac{e\beta_2m^2}{8 \pi^{2}\alpha_2^{2}
}\left[\mbox{ln}\frac{1+\alpha_2} {1-\alpha_2}-2\alpha_2\right].
$$ $$ f_3= \frac{{e\kappa m^2 }}{3 {{\pi
}^2}(4+\tau^{2}+\lambda^2)},\ g_3=-\frac{4 e\kappa\lambda {m^2}}{3
{{\pi }^2}(4+\tau^{2}+\lambda^2)^2}.
$$
Here we used the notations:
    $$
    \alpha_1=\frac{4\kappa}{4-\kappa^2+\tau^2
+\lambda^2}, \ \alpha_2=\frac{4\kappa}{4+\kappa^2+\tau^2
+\lambda^2},\ \ \beta_i=\frac{\lambda \alpha_i}{\kappa},\ i=1,2.
    $$
For the large distance asymptotic we have the following results:
$$
\mbox{F}_i(x_3)\mathrel{\mathop=_ {\x_3\to\infty}}   u_i + \frac{
(-1)^{i+1} e\chi\kappa}{2}-2 f_i +
\frac{v_ie^{-2m|x_3|}}{(4m\pi|x_3|)^{3/2}}
\left\{1+O\left(1/x_3\right)\right\}
$$
The constants $u_i$,$v_i$for $i=1,2$ are written as
$$
u_i = \frac{e m^2}{4\pi^2\alpha_i^2}
\Bigg\{ \frac{ \alpha_i{{\beta_i }^2}} {1-(-1)^i {{\alpha_i }^2}}
 +\sum_{k=1}^2\frac{
 b^{(i)}_k
 \ln \big[-{a^{(i)}_k}-\sqrt{(a_k^{(i)})^ 2-1}\big]}
{\sqrt{(a_k^{(i)})^ 2-1}
 }
\Bigg\},\ v_i=\frac{e\alpha_{i}m^2}{1+\beta_{i}},
$$
where
$$
{a^{(1)}_1}=-\frac{ \beta_1 +i\alpha_1  {\sqrt{{{{\beta_1
}^2}-{\alpha_1 }^2}-1}}}{1+{{\alpha_1 }^2}}, \
{b^{(1)}_1}=\frac{({{\beta_1 }^2}-{{\alpha_1 }^2}) \big( \alpha_1
\beta_1
 -i {\sqrt{{{\beta_1 }^2}-{{\alpha_1 }^2}-1}}\big)}
{2(1+{{\alpha_1 }^2})}, $$ $$ {a^{(1)}_2}=-\frac{ \beta_1
-i\alpha_1 {\sqrt{{\beta_1 }^2-{{{\alpha_1 }^2}-1}}}}{1+{{\alpha_1
}^2}},  \ {b^{(1)}_2}=\frac{({{\beta_1 }^2}-{{\alpha_1 }^2})
\big(\alpha_1 \beta_1  + i {\sqrt{{{\beta_1}^2}- {{\alpha_1
}^2}-1}}\big)}{2(1+{{\alpha_1 }^2})}, $$ $$ {a^{(2)}_1}=-\frac{
\beta_2 +\alpha_2  {\sqrt{{{\alpha_2 }^2}+{{\beta_2
}^2}-1}}}{1-{{\alpha_2 }^2}},\ {b^{(2)}_1}=\frac{({{\alpha_2
}^2}+{{\beta_2 }^2}) \big(\alpha_2 \beta_2
 + {\sqrt{{{\alpha_2 }^2}+{{\beta_2 }^2}-1}}\big)}
{2(1-{{\alpha_2 }^2})}, $$ $$ {a^{(2)}_2}=-\frac{ \beta_2
-\alpha_2 {\sqrt{{{\alpha_2 }^2}+{{\beta_2 }^2}-1}}}{1-{{\alpha_2
}^2}},  \ {b^{(2)}_2}=\frac{({{\alpha_2 }^2}+{{\beta_2 }^2})
\big(\alpha_2 \beta_2  - {\sqrt{{{\alpha_2 }^2}+{{\beta_2
}^2}-1}}\big)}{2(1-{{\alpha_2 }^2})}.
$$
For $u_3$, $v_3$ one obtains the following expressions
$$
u_3=\frac{e\kappa m^2}{3
\pi^{2}(4+\tau^{2}+\lambda^{2})}\left(3-\frac{32
\lambda^2}{(4+\tau^{2}+\lambda^{2})^2}\right), \
v_3=\frac{4e\kappa}{\tau^{2}+(2+\lambda)^{2}}.
$$

\section{Fields generated by simplest defects}
The obtained results demonstrate the non-trivial dependence of the
field $E$ and $H$ on the parameters $\kappa$, $\lambda$, $\tau$.
In the main approximation $E$ and $H$ are linear functions of
$\chi$, and  $E=H=0$, if $\chi=0$, $\kappa=0$. Let us consider the
simplest non-trivial case $\kappa\ne0$, $\chi=\lambda=\tau=0$.

   Asymptotics of $E$ and $H$ for large and small $x_3$ are the following.
  If $q=q^{(1)}$, the defect generates pure electric field $E_3$
   \be
   E_3\mathrel{\mathop\approx_{\x_3\to 0}}\  \frac{e m^2 }{8
\pi^{2}\omega^2 }\left[(1+\omega^2)\mbox{Arctg}(\omega) -
\omega\right]\left(\frac{1}{m^2 x_3^2}-2 \right). \label{5_1}
   \ee
   $$
E_3\mathrel{\mathop\approx_{\x_3\to\infty}} \ -\frac{e m^2}{4
\pi^{2}\omega^{2}}\left[\mbox{Arctg}(\omega)-\omega\right],\
\omega=\frac{4\kappa}{4-\kappa^2}. \label{5_2}
   $$
   For $
q=q^{(2)}$ the field is pure magnetic
   \be H_2\mathrel{\mathop\approx_{\x_3\to\ 0}} \  \frac{em^2}{16 \pi^{2}
\omega^{'2}}\left[(1+\omega^{'2})\mbox{ln}\frac{1+\omega'}{1-\omega'}
-2\omega' \right] \left(\frac{1}{m^2 x_3^2}-2\right). \label{5_3}
   \ee
   $$
   H_2\mathrel{\mathop\approx_{\x_3\to\infty}} \ -\frac{em^2}{8
\pi^{2}\omega^{2'}} \left[\mbox{ln}\frac{1+\omega'}
{1-\omega'}-2\omega'\right], \ \omega'=\frac{4\kappa}{4+\kappa^2}.
   $$
For $q=q^{(3)}$, $E_1=E_2=H_1=H_3=0$, and asymptotics of the
fields $E_3$, $H_2$ are of the form
   \be
E_3\mathrel{\mathop\approx_ {\x_3\to\ 0}}\
H_2\mathrel{\mathop\approx_{\x_3\to\ 0}}\  \frac{e\kappa
m^2}{12\pi^2}\left(\frac{1}{m^2 x_3^2}-2 \right) , \label{5_4}
   \ee
   $$
E_3\mathrel{\mathop\approx_{\x_3\to\ \infty}}\
H_2\mathrel{\mathop\approx_{\x_3\to\ \infty}}\ \frac{e\kappa
m^2}{12\pi^2}.
   $$

In all cases the fields are constant at large distances from the
plane. It is in agreement with well known results for homogeneous
charge and current distributions on the plane in the classical
electrodynamics \cite{Landau,Jackson}. The magnitudes of the
fields are defined by the factors $m^2/16\pi^2$, $m^2/8\pi^2$,
$m^2/4\pi^2$. They are of the order $ 10^{17} \ cm^{-2} \div 4
\cdot 10^{17} \ cm^{-2}$. It is more than the typical surface
density  $ 10^{16} \ cm^{-2}$ of atoms on the boundary of solids.
The field $E(x_3)$ is a monotonic function of $x_3$. Therefore
comparing the large- and small-distance asymptotics (\ref{5_1}),
(\ref{5_2}),  we can obtain an estimation for maximal distance
$d_m$ from the plane for which (\ref{5_1}) can be used: $$
    \frac{1}{m^2d_m^2}\geq 2+ c(\omega),\ \
    c(\omega)\equiv 2\frac{\mbox{Arctg}(\omega)-\omega}
    {\omega -(1+\omega^2)\mbox{Arctg}(\omega) }.
$$ Since $0< c(\omega)\leq 1$, we have $ d_m \sim 1,4 \cdot
10^{-10}cm \div 1,7\cdot 10^{-10}cm$. The ranges of validity
$x_3<d_m$ of asimptotics (\ref{5_3}), (\ref{5_4}) can be found
analogously:\\ $d_m \sim  2,0\cdot 10^{-10}cm\div 2,4\cdot
10^{-10}cm$  for (\ref{5_3}),  and $d_m \sim 1,7\cdot 10^{-10}cm$
for (\ref{5_4}).

Important feature of the fields generated by the considered
defects for $q=q^{(1)}, q^{(2)}$ is that they are singular by
$\kappa=\pm 2$. It means that these values of parameter $\kappa$
are the phase transition points, where in the case $q=q^{(1)}$ the
electrical field $E_3$ is  changed in a sudden way, and in the
case $q=q^{(2)}$ the field $H_2$  become infinite. This phenomena
seems  to be not very surprising since it is similar to the known
supercritical effects induced by perturbation of Dirac field by
attractive $\delta$ potential which causes diving of the ground
state into the Dirac see by finite value of the coupling parameter
\cite{dirac 2,dirac 8,dirac 13}.

The points $\kappa=\pm 2$ are stable in respect to transformation
$\kappa\rightarrow \kappa'= 4/\kappa$, for which
$\omega(\kappa)\rightarrow  \omega(\kappa')= - \omega(\kappa), \
E_{3}(\kappa)\rightarrow E_{3}(\kappa')=-E_{3}(\kappa)$, and
$\omega'(\kappa)\rightarrow \omega'(\kappa')= \omega'(\kappa), \
H_{2}(\kappa)\rightarrow H_{2}(\kappa')=H_{2}(\kappa)$. For
$q=q^{(1)}$ in each point of space the magnitude of  field $E_3$,
considered as a functions of $\kappa$ is restricted: $|E
{(\kappa)}|\leq |E(2)|<\infty$ for all values of $\kappa$.

For the $\lambda= \chi= \tau=0$ the short distance asymptotic is
of the form $ E, H \sim const[1/(m^2x^2)-2]$. In this case the
relative correction of the next to leading term appears to be
independent on parameter $\kappa$ describing specific properties
of the defect on the plane.

\section{Conclusions}
Suggested model describes an infinite plane with homogeneous charge
and current distributions in the framework of QED. Specific
properties of the physical system are characterised by additional
term $S_{def}$ (action of defect) combined with the usual action of
QED into the full action of the model. $S_{def}$ was chosen on the
basis of general principles of QED: locality, gauge invariance and
renormalizability of the theory. It contains 4 parameters which can
be fixed by normalisation conditions.

The calculation of the leading order effects shows that mean field
induced by $S_{def}$ has classical behaviour at large distance from
the plane. Corresponding asymptotic can be used as normalisation
conditions pinpointing interplay of parameters describing the fields
of a plane in classical ED \cite{Landau,Jackson} with ones of
considered model. Along this way one can express four parameters of
$S_{def}$ in terms of the effective charge and current densities and
constants characterising macroscopic properties of material of the
plane.

On short distances the fields $E$, $H$ are singular as functions of
distance $x$ from the defect : $ E, H \sim \mbox{const}/x^2$.
Estimating energy density with usual classical formula
$W=(E^2+H^2)/2$, one obtains its behaviour for
(\ref{5_1}),(\ref{5_3}) and (\ref{5_4}) at short distances as $W(x)\sim
c/x^4 + c'/x^2$, where $c,\ c' $ are finite independent from cut-off
constants.  This singularities representing physical peculiarities
of the model could be predicted with dimensional analysis. It is
similar with one found for scalar field under Dirichlet or Neumann
boundary conditions on a single plate \cite{Romeo-Sahar},  akin
local effects near surfaces can be observed in different geometries
( see, for example, \cite{Fulling,Graham,Milton2}). Our model
predicts dependence of $c$ and $c' $ on parameters of the material
plane.

  Properties of material film described by $S_{def}$
are defined by classical 4-current (given by $S_l(A)$) and particle
density and 4-current of quantum of electron-positron field (given
by $S_{\lambda q}({\bar\psi},\psi)$). Thus, it is assumed that the
film consists of infinitely heavy charged particles, electrons and
positrons uniformly filling the plane. Fluctuation of Dirac field in
QED  generates an effective width of the film. It is of the order
$10^{-10} cm$. In more realistic model one should describe the heavy
charges in the framework of quantum fields theory. Simplest model of
this kind could be QED with additionally proton-antiproton field.
Fluctuations of this field generate anomalous electromagnetic field
at a distance of order $10^{-13} cm$  (because proton is 2000 times
heavier then electron). Thus, they are insufficient for the effects
at a distance of order $10^{-10}cm$ calculated in this paper. The
leading singularity of the electromagnetic field near the plane is
mass independent. Therefore in virtue of opposite charges of
electron and proton, it is canceled for neutral defect with the same
parameters $\lambda, \vec{q}$  for both fermion fields, and at the
distances $x$ less than $10^{-13} cm$ their fluctuation generates
 the fields of the form
$<E>\sim (M-m)C_e/|x|+(M^2-m^2)C'_e$, $<H>\sim (M-m)C_h/|x|+
(M^2-m^2)C'_h$, where $M$ $(m)$ is the mass of proton (electron),
$C_e,\ C'_e,\ C_h,\ C'_h$ are functions of parameters $\lambda,
\vec{q}$, and $C_e =C _h =0$ if $\lambda=0$.

   In the model of homogeneous defect the
the surface energy density is infinite because of ultra-violet
divergences, and needs a regularization with cut-off impulse
$\Lambda$ forming effectively a discreet lattice structure with
characteristic scale $\sim \Lambda^{-1}$ in the coordinate space. It
can be considered as a model of film consisting of atoms. Analysis
of the function $F(x_3,\Lambda)$ used for calculations of the fields
${\cal F}_{\mu\nu} (x_3)$ show that fields near the plane for
discreet defect ${\cal F}_{\mu\nu} (x_3)$  can be described if we
replace in results for continuous defect the singularities of ${\cal
F}_{\mu\nu} (x_3)$ generated by fermion field of mass $m_f$ as
follows: $1/x_3^2 \rightarrow f_2(x_3,\mu)=(1-(1+\mu
x)\exp{-\mu|x_3|})/x_3^2$, $1/|x_3| \rightarrow f_1(x_3,\mu)=(1
-\exp{-\mu |x_3|})/|x_3|$ where $\mu=\mbox{min}\{\Lambda, m_f\}$.
Such approximations are valid for distances smaller as $\sim 1/\mu$.
Thus, for neutral discreet defect with $\Lambda^{-1}\sim
10^{-8}-10^{-7} cm $ the behaviour of the fields ${\cal F}_{\mu\nu}
(x_3)$ on the distances $\sim \Lambda^{-1} $ can be approximately
described by main terms of short distance asymptotic of the
homogeneous model. It is not surprising since there is no sharp
cutoff dependence in the model of renormalizable quantum field
theory.

    In our paper we have restricted ourself with the simple problem
to calculate the mean electromagnetic field generated by
perturbation of QED vacuum by infinite plane film. It is important
to note that a finite physical observable is extracted for a system
of just one isolated plane in distinction to ordinary Casimir
effect. Obtained formulas predict dependence of Casimir-Polder
forces on charge and current densities which can be proved
experimentally. In the the critical point $\kappa=\pm 2$ the
magnetic field becomes infinite, and in the vicinity of critical
point it must be observable near the defect surface with overcoming
exponential suppression from masses of fermions. Other way to test
the proposed model is the usage of obtained results for calculation
of some physical feature of the film (for example, the spectrum of
bound states of charged particle in anomalous electric potential of
neutral defect) which can be investigated experimentally. In the
framework of the model one can study  scattering of electrons and
photons on the defect and obtain results for experimental verifying.

   We hope that the proposed approach can provide
a deeper insight into the nature of quantum phenomena of interaction
of macroscopic bodies with QED fields.

\section*{Acknowledgements}
V.N.~Markov was supported in part by
DAAD , Dynasty Foundation and personal grant of the governor of St.Petersburg.
V.N.~Markov is also indebted to Prof. H.R.~Petry for
his hospitality during the stay in Bonn and to V.V. Vereshagin and
M. Vyazovsky for useful discussions. The work of Yu.M. Pis'mak was
supported in part by Russian Foundation   for  Basic   Research
(Grant  No 03-01-00837), and Nordic Grant for Network Cooperation
with the Baltic Countries and Northwest Russia No FIN-6/2002.
Yu.M.~Pis'mak is grateful to H.W.~Diehl for fruitful discussions.


\end{document}